# Hollow-core fibers with reduced surface roughness and ultralow loss in the short-wavelength range

Jonas H. Osório[1], Foued Amrani[1, 2], Frédéric Delahaye[1, 2], Ali Dhaybi[1], Kostiantyn Vasko[1], Fabio Giovanardi[3], Damien Vandembroucq[4], Gilles Tessier[5], Luca Vincetti[3], Benoît Debord[1, 2], Frédéric Gérôme[1, 2], Fetah Benabid[1, 2, *]

[1]GPPMM Group, XLIM Institute, CNRS UMR 7252, University of Limoges, Limoges, 87060, France
[2]GLOphotonics, 123 Avenue Albert Thomas, Limoges, 87060, France
[3]Departament of Engineering "Enzo Ferrari", University of Modena and Reggio Emilia, Modena, 41125, Italy
[4]PMMH, CNRS UMR 7636, ESPCI Paris, PSL University, Sorbonne University, Paris Cité University, Paris, 75005, France
[5]Vision Institute, CNRS UMR 7210, Sorbonne University, Paris, 75012, France

**Abstract:** While optical fibers display excellent performances in the infrared, visible and ultraviolet ranges remain poorly addressed by them. Obtaining better fibers for the short-wavelength range has been restricted, in all fiber optics, by scattering processes. In hollow-core fibers, the scattering loss arises from the core roughness and represents the limiting factor in reducing their loss regardless of the fiber cladding confinement power. To attain fibers performing at short wavelengths, it is paramount developing means to minimize the height variations on the fiber microstructure boundaries. Here, we report on the reduction of the core surface roughness of hollow-core fibers by modifying their fabrication technique. In the novel process proposed herein, counter directional gas fluxes are applied within the fiber holes during fabrication to attain an increased shear rate on its microstructure. The effect of the process on the surface roughness has been quantified by optical profilometry and the results showed that the root-mean-square surface roughness has been reduced from 0.40 nm to 0.15 nm. The improvement in the fiber core surface quality entailed fibers with ultralow loss in the short-wavelength range. We report on fibers with record loss values as low as 50 dB/km at 290 nm, 9.7 dB/km at 369 nm, 5.0 dB/km at 480 nm, and 1.8 dB/km at 719 nm. The results reveal this new approach as a promising path for the development of hollow-core fibers guiding at short wavelengths with loss that can potentially be orders of magnitude lower than the ones achievable with their silica-core counterparts.

## Introduction

Hollow-core photonic crystal fibers (HCPCFs) approach their 30th anniversary [1]. Their remarkable performances, demonstrated in both fundamental and applied fields, justify the great efforts devoted by the HCPCF community toward a better understanding of their properties, the optimization of their designs and fabrication processes, as well as the consolidation of the application fields. Historically, photonic bandgap (PBG) fibers [1] were the first HCPCFs to appear as an alternative to overcome the fundamental constraint imposed by the Rayleigh scattering limit of glass and attain ultralow loss, particularly in the visible and ultraviolet spectral ranges, where silica attenuation dramatically increases. PBG fibers, however, have been dismissed as eligible candidates to surpass solid-core silica fibers loss levels due to limitations such as the strong core-cladding optical overlap, the presence of surface modes, and their core surface roughness. Additionally, having PBG fibers guiding at short wavelengths (< 1 μm) requires smaller cladding pitches, which further complicates PBG fibers' fabrication processes. The lowest loss reported for PBG fibers is 1.2 dB/km around 1600 nm [2]. At shorter wavelengths, the minimum attenuation is much higher: 870 dB/km at 557 nm [3].

Alternatively, inhibited-coupling (IC) guiding fibers [4] have been proved to exhibit confinement loss (CL) figures comparable to PBG fibers but with the outstanding difference that the core-cladding optical overlap is several orders of magnitude smaller. Moreover, as IC fibers work on the large pitch regime, greater cladding pitches can be used to operate at short wavelengths. Furthermore, the criteria for IC guidance cause the fiber CL to be strongly dependent on the core contour shape. It inspired, for example, the proposal of the hypocycloid core contour (negative curvature) in 2010 [5, 6], which provided a dramatic reduction of the loss in IC fibers.

The current state-of-the-art loss levels in IC fibers are set by the surface-roughness scattering loss (SSL) for fibers guiding at short wavelengths (< 1 μm) and by the fiber design for fibers guiding at longer ones [7, 8]. Concerning the latter, alternative cladding design to kagome and single-ring tubular lattice (SR-TL) HCPCF has entailed ultralow loss figures in the infrared range [9-12]. On the other hand, for shorter wavelengths, SR-TL HCPCFs have provided loss values of 7.7 dB/km at 750 nm and 13.8 dB/km at 539 nm [7, 13]. Nested-tubes HCPCFs, in turn, display loss figures of 1.4 dB/km at 862 nm and 2.85 dB/km at 660 nm [11]. Finally, conjoined-tubes HCPCFs show the figures of 3.8 dB/km at 680 nm and 4.9 dB/km at 558 nm [14]. In the UV range, loss figures values of 100 dB/km at 218 nm and 130 dB/km at 300 nm have been measured [15, 16].

In this framework, although improvements in the cladding design of IC guiding fibers have allowed decreasing the loss figures in the infrared range, obtaining fibers for the visible and ultraviolet range remains a more challenging task due to the SSL limitation. Indeed, SSL is set by the core surface roughness, which arises from thermal surface capillary waves (SCW) that are frozen during the fiber draw [2]. SSL assumes the form of $\alpha_{SSL} = \eta \times F \times (\lambda/\lambda_0)^{-3}$ (where $F$ is the core mode overlap with the core contour, $\lambda$ is the wavelength, and $\lambda_0$ is a calibrating constant) and scales quadratically with the surface roughness root-mean-square (rms) height, i.e. $\alpha_{SSL} \propto h_{rms}^2$ [2]. Indeed, the factor $\eta$ in $\alpha_{SSL}$ formula relates to the surface quality, being proportional to $h_{rms}^2$. Reducing SSL, therefore, implies controlling and/or reducing the roughness height. In this context, despite recent results show that adjusting the drawing stress during the fiber fabrication process of fibers diminishes the roughness along the drawing direction [17], no solution has been demonstrated so far on how to mitigate SSL-dominated scenarios in HCPCFs whatsoever.

Here, we report that, by revisiting the HCPCFs fabrication technique, reduction of the rms roughness of the core surfaces of HCPCFs is possible. Our innovative technique incorporates the concept that shear flow can attenuate capillary waves [18, 19] in the fiber drawing process. Applying counter directional glass fluxes within the fiber holes during the fiber fabrication provides an increased shear rate on the membranes of SR-TL HCPCFs. In our investigation, we studied two sets of fibers, the first one produced by using the standard HCPCF fabrication method and the second one by using this innovative technique. Optical profilometry measurements showed that the rms roughness was reduced from 0.40 nm down to 0.15 nm via the utilization of the technique proposed herein. The reduction of the core surface roughness allowed to obtain fibers with loss figures as low as 50 dB/km at 290 nm, 9.7 dB/km at 369 nm, 5.0 dB/km at 480 nm, and 1.8 dB/km at 719 nm. We believe that our results provide a new framework for the development of optical fibers transmitting light in the visible and ultraviolet ranges, and open exciting prospects in UV-photonics.

**Results**

**Surface capillary waves within HCPCFs' scenario**

During HCPCF fabrication, the structured glass preforms undergo heating inside a high-temperature furnace. The heating process entails melting of the preform, which allows, by suitably pulling the fiber and pressurizing its internal microstructure, to successfully fabricate the desired HCPCF architecture. Within the heating process context, the dynamics of surface capillary waves (SCW) is established as a result of two competing effects, namely the thermal noise, which is prone to ruffle the surface, and the interface tension, which tends to attenuate the SCW oscillations. Indeed, the SCW dynamics have been proved to govern the surface roughness of the glass surfaces in HCPCFs, as they are abruptly frozen at the glass transition temperature, $T_G$ [2]. Under the SCW framework, the Fourier spectrum of the surface height profiles, which is typically assessed via their power spectral density (PSD) functions calculated over 1D surface profiles, is expected to display a $1/f$ behavior, where $f$ is the spatial frequency. The $1/f$-trend behavior of PSD functions accounted from HCPCFs core surface profiles has been observed in previous works [2, 19-21]. For low spatial frequency values (< $5 \times 10^{-2}$ µm$^{-1}$), however, deviation from the $1/f$-trend has been observed, and PSD functions following a $1/f^3$-trend have been detected [21]. The latter is likely to be due to boundary conditions imposed by the heat zone length of the drawing furnace.

As mentioned before, we here develop a new technique for reducing the roughness of HCPCFs core surfaces to attain ultralow loss figures in the short-wavelength range. In our study, we employ SR-TL HCPCFs due to their potential to provide ultralow loss in the visible and ultraviolet spectral ranges allied to a simple fiber structure from a fabrication viewpoint. Fig. 1a presents the cross-section of one of the SR-TL HCPCFs used in our investigations, which has been fabricated by using standard HCPCF fabrication methods. To assess the quality of the core surfaces in the fiber, optical profilometry and atomic force microscopy (AFM) have been employed. Fig. 1b shows typical height profiles (Δh) along the fiber axis measured by using an AFM (top plot) and a picometer-sensitivity optical profilometer (bottom plot) [20, 21]. Fig. 1c, in turn, presents a representative AFM-measured surface profile taken for a 6 μm × 6 μm surface on the fiber core surface. The optical profilometer and AFM data allow obtaining the PSD plot shown in Fig. 1d. Optical profilometry is limited to low spatial frequencies (a few $10^{-1}$ μm$^{-1}$) by diffraction and probe spot spacing. Conversely, the scanning ranges limit AFM to higher frequencies, between 1 and a few tens of μm$^{-1}$. The combination of the two measurement techniques provides a spatial frequency interval ranging from $1 \times 10^{-2}$ μm$^{-1}$ to 20 μm$^{-1}$. For comparison, we include in the plot a $1/f$ trend calculated from Eq. (1) given below:

$$|H_{SCW}(f)|^2 = \frac{k_B T_G}{2\pi \gamma f} \quad (1)$$

where $k_B$ is the Boltzmann constant and $\gamma$ is the surface tension –, which stands for the PSD expected from 1D surface profiles under SCW scenario. Here, we used $T_G/\gamma$ = 2000. It is seen that the PSD trend deviates from the $1/f$ behavior for $f < 1 \times 10^{-2}$ μm$^{-1}$, similarly to what has been found in recent investigations [20, 21]. While the $1/f$ behavior directly results from the freezing of surface capillary waves, the deviation of it at lower frequencies can be ascribed to boundary size effects.

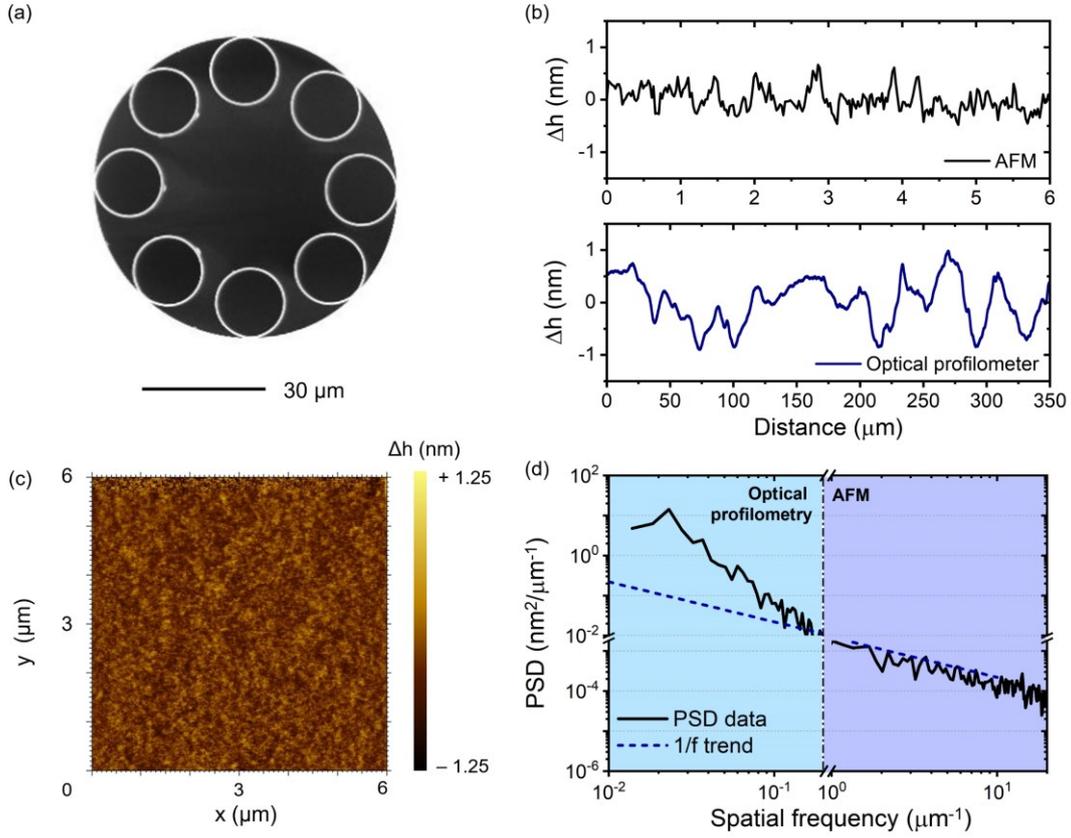

**Figure 1. (a)** Cross-section of a SR-TL HCPCF. **(b)** Typical surface profiles for a fiber drawn by using typical HCPCF fabrication methods measured by AFM (top plot) and optical profilometry (bottom plot). **(c)** Typical AFM measurement result for a fiber drawn by using typical HCPCF fabrication methods. **(d)** PSD plot accounted from optical profilometry and AFM measurements.

**Shear stress as means to structure the surface roughness profile**

The inherent roughness of glass surfaces arises from fluctuations of capillary waves, which, as discussed in the last section, result from the interplay between the thermal noise and the glass interface tension [22]. In the absence of shear, the rms height of these frozen fluctuations amounts to $\sqrt{\frac{k_B T_G}{\gamma}} \approx 0.4 nm$, where $T_G \approx 1500\ K$ and $\gamma \approx 0.3\ J/m^2$ typically. This rms height level is, hence, referenced as thermodynamic equilibrium surface roughness (TESR).

In 2006, Derks *et al.* [19] concluded that shear can suppress capillary waves and make interfaces smoother. To describe this effect, the authors introduced the concept of an effective interfacial tension, $\gamma_{eff}$, to be assigned to sheared systems, which increases with the shear rate. According to the model proposed in [19], the effective interfacial tension can be expressed by Eq. (2), where $\gamma_0$ is the interfacial tension at zero shear and $\Phi(\Gamma)$ is a positive function that grows with increasing shear rate $\Gamma$.

$$\gamma_{eff}(\Gamma) = \gamma_0 + \Phi(\Gamma) \tag{2}$$

As the mean square roughness, $\langle h^2 \rangle$, is inversely proportional to the interfacial tension, Derks *et al.* stated that the amplitude of the capillary waves can be expressed as Eq. (3), where $\langle h^2 \rangle(\Gamma = 0)$ is the mean square roughness under zero shear [19]. Therefore, the mean squared height of sheared surfaces is expected to be lower than the value for not sheared ones.

$$\langle h^2 \rangle(\Gamma) = \left[\frac{\gamma_0}{\gamma_0 + \Phi(\Gamma)}\right] \langle h^2 \rangle(\Gamma = 0) \tag{3}$$

Such a reduction of interfacial fluctuations has also been studied by Thiébaud *et al.* [18] and Smith *et al.* [23, 24]. Particularly, Smith *et al.* [23, 24] performed simulations on the dynamics of laterally driven surfaces and concluded that the existence of shear implies the reduction of the interfacial width. They affirmed that shear acts as an effective confinement force in the system which can suppress the interfacial capillary wave fluctuations. Similarly to [19], they argued that defining a "nonequilibrium surface tension" to theoretically fit the behavior of the height-height correlation functions of the sheared surfaces entails an increase of such tension as the system is more strongly driven.

Within the context of optical fibers, the reduction of the roughness of sheared glass surfaces has been assessed by Bresson *et al.* on glass tubes [17]. In the latter, by evaluating the surface profiles of the glass tubes with typical diameters of 220 µm and thicknesses of 15 µm, the authors identified that the surface roughness levels can be lowered by the fiber drawing process thanks to the attenuation of the surface capillary waves along the drawing direction. In this context, they detected an anisotropic behavior of the height correlations in the fabricated fibers and showed that the glass surfaces can retain a structural signature of the direction of the flow that took place during the fiber fabrication.

The parameter chosen in [17] for studying the surface capillary waves attenuation, the tension experienced by the fiber during fabrication, results however from several draw parameters such as furnace temperature, draw speeds, and the dimensions of fiber to be drawn and of its preform. Hence, although it had allowed the authors in [17] to correlate the lessening of the surface roughness levels in the studied glass tubes with the drawing process, such a parameter does not present itself as the most adequate indicator of flow establishment inside HCPCFs for our investigation, whose aim is to study the reduction of the surface roughness levels inside SR-TL HCPCFs.

Thus, in order to establish a flow inside the microstructure of SR-TL HCPCFs, we have worked with the application of opposite pressure gradients inside the preforms during fiber fabrication. Hagen-Poiseuille's law in fluid dynamics and the definition of the shear rate inside a tube apprises us that the shear rate can be correlated with the pressure gradients experienced by the preform components during the fiber drawing. In our investigation, we minimized the drawing down ratio (*i.e.*, the ratio between the preform outer diameter and the fiber outer diameter) and judiciously designed the preform and fiber dimensions to accommodate the shear flow conditions. Details on the fabrication procedures will be provided in the following.

**Surface roughness reduction in SR-TL HCPCFs**

The usual technique for drawing SR-TL HCPCFs employs the pressurization of core and cladding regions, so adequate dimensions are achieved on the fiber cross-section. Current techniques, thus, apply co-directional gas flows to the fiber structure during its fabrication so the desired dimensions of the cladding elements can be obtained. Differently, here we propose the utilization of counter-directional gas flows within the fiber microstructure during the fiber draw for adding shear to the microstructure's glass membranes and, hence, attaining smoother core surfaces. In our technique, vacuum is applied to the fiber core, while the ring of tubes in the SR-TL HCPCF structure is pressurized (Fig. 2a). Fig. 2b presents a diagram for the interfaces between the surfaces of the cladding tubes, the fiber core, and the internal region of the cladding tubes.

To demonstrate the reduction of the core surface roughness via the utilization of the approach proposed herein, we systematically studied two sets of SR-TL HCPCFs, both fabricated by using the stack-and-draw technique. The first group (G#1, composed of five different SR-TL HCPCFs fabricated in independent fiber draws, with cladding tubes thickness ranging from 230 nm to 580 nm) was produced by using the standard fabrication method for HCPCFs. Fibers in the second group (G#2, composed of nine different SR-TL HCPCFs fabricated in independent fiber draws, with cladding tubes thickness ranging from 300 nm to 1220 nm), instead, were produced by the innovative technique reported herein, *i.e.*, during fiber fabrication, vacuum was applied to the fiber core while a gas flow was used to inflate the cladding tubes.

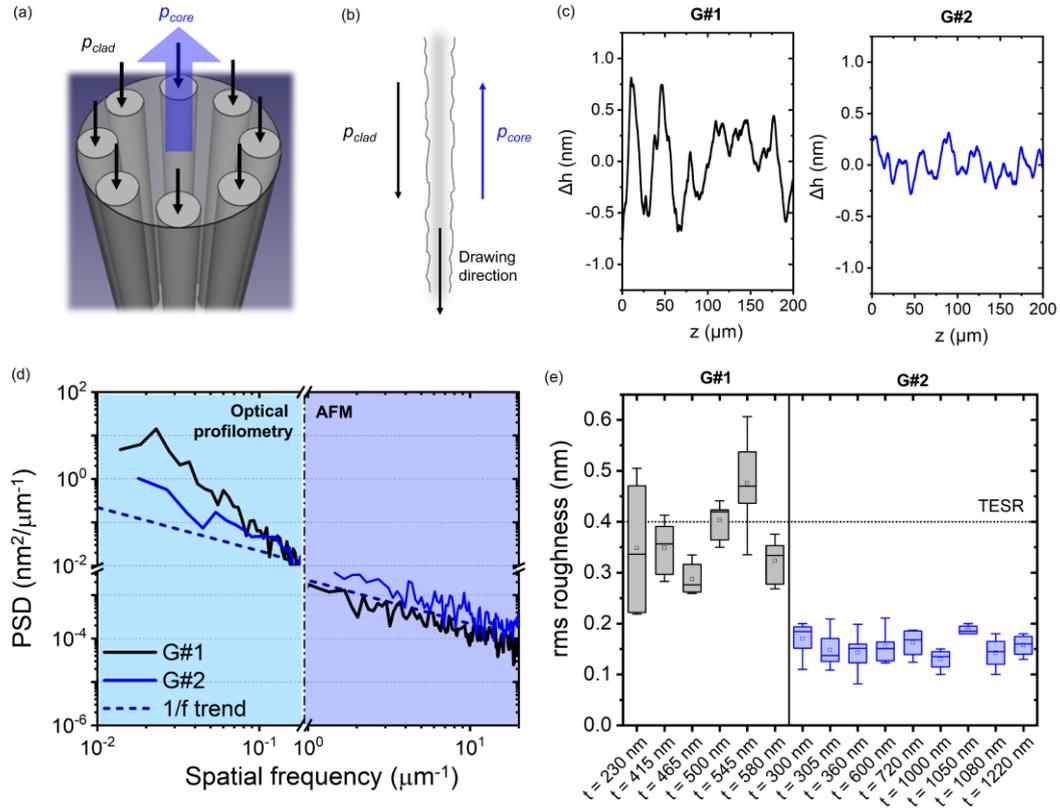

**Figure 2. (a)** Diagram for the pressure application in the fiber preform during the innovative process proposed in this manuscript; $p_{core}$: pressure in the fiber core; $p_{clad}$: pressure in the cladding tubes. **(b)** Diagram of the interfaces between the surfaces of the cladding tubes, the tubes internal region and the fiber core. **(c)** Typical roughness profiles along the fiber axis and **(d)** PSD for fibers in G#1 and G#2. **(e)** Rms roughness as a function of the thickness of the cladding tubes, t, for fibers in G#1 and G#2. TESR: thermodynamic equilibrium surface roughness.

The core roughness profiles of the fibers in G#1 and G#2 have been experimentally measured by using a picometer-resolution optical profilometer [20, 21]. Fig. 2c shows typical surface profiles for fibers in G#1 (left-hand side) and G#2 (right-hand side) measured in the optical profilometry experiments. It is seen that fibers in G#2 present reduced peak-to-peak roughness values compared with fibers in G#1 (for fibers in G#1, peak-to-peak values are around 1.5 nm, and, for fibers in G#2, around 0.5 nm). Moreover, Fig. 2d exposes typical PSD traces for fibers in G#1 and G#2. Noteworthily, the new fabrication methods reported herein entail a reduction of the PSD values at spatial frequencies lower than $10^{-1}$ μm$^{-1}$, which, in turn, readily impacts the rms roughness of the fibers.

In this context, Fig. 1e shows box charts for the rms roughness values, accounted from the optical profilometry measurements, for fibers in G#1 (left-hand side) and G#2 (right-hand side) as a function of the thickness of the cladding tubes (t). Values in Fig. 2e were obtained from the rms roughness values measured in several scans (with typical lengths of 200 μm) for the fiber samples in G#1 and G#2. It is seen that fibers in G#1 have rms roughness values that oscillate around 0.40 nm while fibers in G#2 have rms roughness values around 0.15 nm. Indeed, the application of counter directional gas flows during the fiber drawing allowed to enhance the quality of the core surface of the HCPCFs by a factor of 2.7 and attain fibers with core surface roughness lower than the thermodynamic equilibrium level, whose rms value amounts to $\sqrt{\frac{k_B T_G}{\gamma}} \approx 0.4 nm$.

**Ultralow loss in SR-TL HCPCF in the short-wavelength range**

As described in the last section, the use of counter-directional gas flows within the fiber structure during the fiber drawing has entailed fibers with smoother core surfaces. A direct effect of the reduction of the roughness levels of the core surfaces is the lowering of the SSL in the fabricated fibers. In this framework, we report on SR-TL HCPCFs with ultralow loss in the short-wavelength range (i.e. λ < 1 µm). Prior to showing the new records achieved, we firstly develop on the impact of SSL reduction on HCPCF loss trends.

The study of the loss spectrum of SR-TL HCPCFs allows determining empirical scaling laws governing the attenuation trends in such fiber structures. Thus, based on simulated loss spectra for representative structures, it has been previously reported that the CL in SR-TL HCPCF decreases for shorter wavelengths following a $\lambda^{4.5}$ trend [25]. Thus, by considering that the total loss (TL) in the fibers is given by TL = SSL + CL, one can calculate the TL trends based on such scaling law.

Fig. 3a displays the TL trends (together with the SSL and CL ones) which have been calculated via the above-mentioned scaling laws and by assuming a fiber structure with a core diameter of 40 µm. The different TL curves in Fig. 3a have been obtained by plugging different values of $\eta$ in the SSL scaling law, namely $\eta = 2100$ (black curve) and $\eta = 300$ (blue curve). The latter values have been chosen to illustrate fibers with greater ($\eta = 2100$, estimated by adjusting it to the loss of fibers with thermodynamic equilibrium surface roughness [2]) and lower ($\eta = 300$, corresponding to a fiber with rms core surface roughness 2.7 times lower than the thermodynamic equilibrium surface roughness, as SSL is proportional to the square of the rms roughness of the core surfaces [2]) contributions of the SSL to the TL. Observation of Fig. 3a readily informs that the TL curves display a turning point wavelength (TL curve minimum) for which the contribution of SSL to the TL becomes dominant. Moreover, one can observe that the turning point in the TL curve (identified by the vertical lines in Fig. 3a) is pushed towards shorter wavelengths when one assumes smaller $\eta$ (*i.e.*, when we consider fibers with lower SSL), as indicated by the arrow in Fig. 3a.

The former analysis can be used to further substantiate the impact of the core surface roughness improvement reported in the last section on the loss of the fibers. Thus, Fig. 3b presents the loss spectra of representative fibers in G#1 and G#2 with similar cladding tube thicknesses (t ~ 0.6 µm). By observing the loss spectrum of G#1 fiber (black curve; core diameter 41 µm), one sees that there is a considerable loss increase for wavelengths lower than 600 nm due to SSL impact. Otherwise, the results for G#2 fiber (blue curve; core diameter 27 µm) show that a decreasing loss trend is maintained even for wavelengths lower than 600 nm (the difference between the loss values for the band between 600 nm and 1000 nm is due to the different core diameters of the fibers considered in this analysis). The distinction between the loss behaviors of fibers in G#1 and G#2 is, therefore, the optical manifestation of the SSL reduction in G#2 fibers.

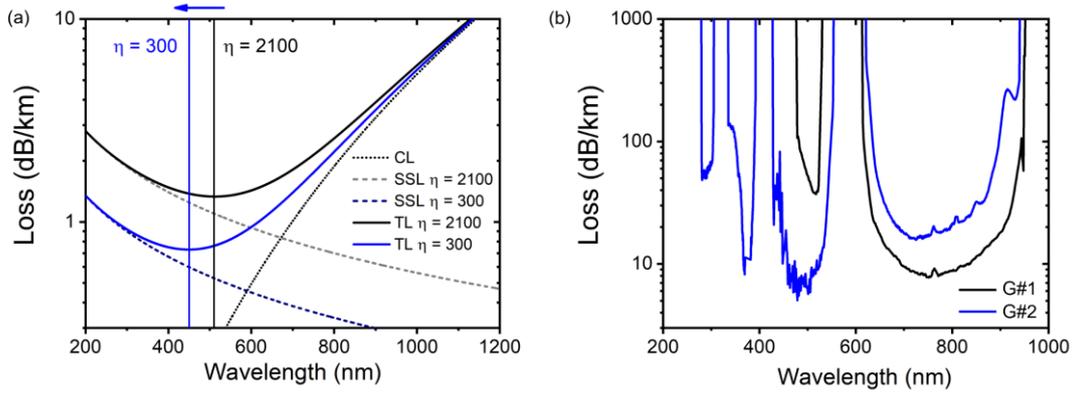

**Figure 3. (a)** CL, SSL, and TL trends calculated from scaling loss by considering two different values for η. **(b)** Representative measured loss spectra for fibers in G#1 and G#2 showing the different loss trends in the short-wavelength range.

Finally, we report on record attenuation values for fibers guiding at short wavelengths. Fig. 4a and Fig. 4b exhibit the cross-sections of the fibers we report here (referenced as Fiber A and Fiber B, both fabricated by following the innovative fabrication methods reported herein) and their corresponding measured loss. Fiber A and Fiber B are SR-TL HCPCFs with cladding formed by a set of 8 untouching cladding tubes. The minimum loss figures of Fiber A are 50.0 dB/km at 290 nm, 9.7 dB/km at 369 nm, and 5.0 dB/km at 480 nm. In turn, the minimum attenuation values for Fiber B are 0.9 dB/km at 558 nm and 1.8 dB/km at 719 nm. Table 1 displays the fibers' geometrical parameters.

Fig. 4c contextualizes the results on Fiber A and Fiber B within the IC HCPCF framework in the wavelength interval between 250 nm and 900 nm. Additionally, Fig. 4c presents the silica Rayleigh scattering limit trend (SRSL). Noteworthily, Fiber A and Fiber B loss figures stand for new record low-loss values in the short-wavelength range and lie under the SRSL, which is a fundamental limitation that hinders the reduction of attenuation values of silica-core fibers. Additionally, it is remarkable that such ultralow loss figures have been attained by using a fiber architecture as simple as SR-TL HCPCFs without the need for other more complex IC HCPCF structures.

**Table 1.** The geometrical parameters of Fiber A and Fiber B ($D_{core}$: diameter of the core; t: tubes' thickness; $g$: gap between the cladding tubes; $D_{tubes}$: diameter of the cladding tubes).

| Fiber | $D_{core}$ (µm) | t (µm) | $g$ (µm) | $D_{tubes}$ (µm) |
|---|---|---|---|---|
| A | 27 | 0.6 | 2.1 – 4.7 | 11 |
| B | 42 | 0.9 | 3.6 – 5.2 | 18 |

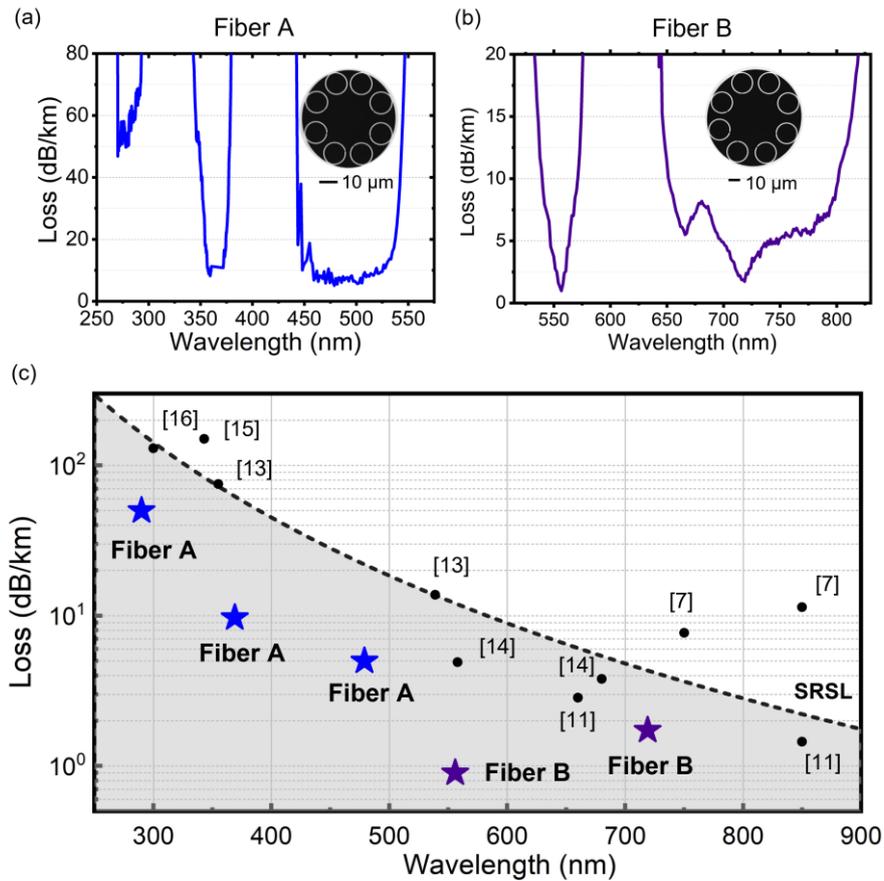

**Figure 4.** Loss measurement results for **(a)** Fiber A and **(b)** Fiber B and the fibers' cross sections. **(c)** IC HCPCF state-of-the-art framework and the SRSL trend.

## Discussion

Current developments in visible and ultraviolet photonics call for fibers able to transmit such wavelengths with low loss. Improvements in the loss displayed by solid-core fibers, however, tend to be marginal as they are fundamentally restricted by silica properties. HCPCFs, on the other hand, emerge as promising alternatives to circumvent silica-core fibers limitations. However, the performances of HCPCF working at short wavelengths have been hitherto limited by scattering processes arising from the roughness of the fiber core surfaces.

In this manuscript, we reported on HCPCFs with reduced core surface roughness, which have been obtained by modifying these fibers' usual fabrication techniques. Here, we proposed the use of counter-directional gas flows during the fiber draw to add shear to the fiber silica membranes and, thus, achieve smoother core surfaces. We showed that the rms roughness values were reduced from 0.40 nm to 0.15 nm via the application of the novel process proposed herein.

The amelioration of the core surface quality – and, therefore, the reduction of SSL – allowed attaining fibers with new record low loss figures in the short-wavelength range. Remarkably, the new record loss figures have been accomplished by using a HCPCF structure as simple as the SR-TL HCPCF design. We understand that the results presented in this paper identify a step-change path for the development of optical fibers operating at short wavelengths. We envisage that our achievements will motivate new research efforts towards further reduction of core surface roughness in HCPCFs and, hence, towards the demonstration of even lower attenuation values, especially in the visible and ultraviolet spectral ranges.

**Materials and methods**

**Optical profilometry:** picometer-resolution optical profilometry has been used to assess the height profiles of the HCPCF. The profilometer working principle relies on the reflection and interference of two polarization-modulated laser beams that impinge on spatially-separated sites of the tested sample [20, 21]. The optical profilometry measurements have been performed by immersing the fiber under test and by filling the cladding tubes with index-matching liquid to avoid unwelcome reflections.

**AFM measurements:** a commercial atomic-force microscope has been used to characterize the HCPCF core surfaces at high spatial frequency values. For the AFM measurements, the fiber has been angle-cleaved so a clean and debris-free region could be obtained for the realization of the tests.

**Loss measurements:** for wavelengths larger than 400 nm, the attenuation values were obtained from cutback measurements using light from a supercontinuum source and an optical spectrum analyzer. When accounting for the loss at wavelengths shorter than 400 nm, a plasma lamp was used as the light source, and the transmitted signal was measured in a spectrometer. In the setup for loss measurements in the UV range, one placed a flip mirror in-between the fiber end and the spectrometer so to allow correct positioning of the fiber output by observing its image in a CCD camera. When the positioning of the fiber end was assured to be correct, the mirror was flipped, and the spectrometer measured the transmission signal.

**Data availability:** The data that support the findings of this study are available from the corresponding author upon reasonable request.

**Acknowledgment.** This research has been funded through PIA program (grant 4F), OzoneFinder project, and la région Nouvelle Aquitaine.

**Conflict of interests:** The authors declare that they have no conflict of interest.

**Contributions:** F.B. directed the work. J.H.O., F. A., F.D., A. D., B.D., and F.Ge. worked on fiber fabrication. J.H.O. performed the optical characterization measurements. J.H.O., A. D., and G.T. performed the profilometry measurements. J.H.O., F.Ge., and F.B. wrote the paper. K. V., F. Gi., and L.V. worked on simulations. J.H.O., G. T., D.

V., and F. B. assessed the surface roughness results. All the authors discussed the results and reviewed the manuscript.